\documentclass[conference]{IEEEtran}
\usepackage[latin9]{inputenc}
\usepackage{amsmath}
\usepackage{amssymb}
\usepackage{graphicx}
\usepackage{booktabs}
\usepackage{multirow}
\usepackage{url}
\renewcommand{\arraystretch}{0.85}
\usepackage{xcolor}
\usepackage{cite}
\usepackage{balance}
\usepackage{tikz}
\usetikzlibrary{arrows.meta,positioning,shapes.geometric,fit,backgrounds,calc}

\makeatletter
\def\ps@IEEEtitlepagestyle{
  \def\@oddfoot{\mycopyrightnotice}
  \def\@evenfoot{}
}
\def\mycopyrightnotice{
  {\footnotesize 979-8-3195-0323-7/26/\$31.00 \copyright2026 IEEE\hfill}
  \gdef\mycopyrightnotice{}
}

\@ifundefined{showcaptionsetup}{}{
 \PassOptionsToPackage{caption=false}{subfig}}
\usepackage{subfig}
\makeatother

\usepackage{eso-pic}
\newcommand\AtPageUpperMyright[1]{\AtPageUpperLeft{
 \put(\LenToUnit{0.5\paperwidth},\LenToUnit{-1cm}){
     \parbox{0.5\textwidth}{\raggedleft\fontsize{9}{11}\selectfont #1}}
 }}
\newcommand{\conf}[1]{
\AddToShipoutPictureBG*{
\AtPageUpperMyright{#1}
}
}

\newif\ifnbfig\nbfigfalse 

\begin{document}

\title{Quantifying the Intrusiveness of Retrofitting Retrieval-Augmented Generation onto Legacy Software Systems}
\conf{2026 IEEE 8th International Conference on Sustainable Technologies for Industry 5.0 (STI 2026), 03-04 December, Dhaka}
\author{
  \IEEEauthorblockN{
    \textbf{K.M.Tahlil Mahfuz Faruk\textsuperscript{1}}, \textbf{Rakib Abdullah\textsuperscript{2}}
  }
  \IEEEauthorblockA{
    \textsuperscript{1}\textit{Department of Software Engineering, Green University of Bangladesh}\\
    \textsuperscript{2}\textit{Department of AI and Data Science, Green University of Bangladesh}\\
    Email: \textsuperscript{1}tahlil@swe.green.edu.bd, 
    \textsuperscript{2}rakib@ads.green.edu.bd
  }
}


\maketitle

\begin{abstract}
Enterprises increasingly add retrieval-augmented generation (RAG) to existing REST and monolithic systems,
usually by writing an ad hoc HTTP call into a handler, and have no way to quantify how intrusive that choice is.
We present (1) a taxonomy of three retrofit patterns, \emph{sidecar}, \emph{SDK-embed} and
\emph{naive-invasive}; (2) the Legacy Intrusion Score (LIS), three code sub-metrics (churn, breadth, coupling),
and an extension, LIS-X, with three operational sub-metrics measured on a live system (run-time overhead, fault
propagation and silent feature degradation); (3) a static evaluation on five legacy systems, two of them real
open-source codebases fetched from their repositories at pinned commits; and (4) a live case study that deploys
all three patterns as separate OS processes. The code-level ranking sidecar~$<$~SDK-embed~$<$~naive-invasive holds
in all five systems and under 2,000 random weightings, but it follows from component-wise dominance of the
sub-metrics, so it validates the metric rather than testing a hypothesis. In the live study, outcomes are not fixed
by definition. The sidecar adds no detectable latency to the legacy endpoint (0.07~ms, 95\% CI $[-0.10,\,0.24]$),
whereas SDK-embed and naive-invasive add 569 and 475~ms and are not distinguishable from each other
($p{=}0.62$). Both pass a 2~s dependency delay on to legacy requests, and when the RAG layer
fails fast they keep returning successful responses that no longer contain the AI field. Load on the RAG
layer slows the legacy service by 1.51--1.57$\times$ on a shared host, and sidecar staleness averages 911~ms with a
2~s polling interval. Retrieval hit@1 is 0.95; three hosted LLMs answer all 12 test questions correctly and a local model 9/12,
and the four backends are swapped by changing one request field. All dynamic results come from one host and one system.
\end{abstract}


\section{Introduction}
\label{sec:intro}

Retrieval-augmented generation (RAG)~\cite{lewis2020rag,gao2024ragsurvey} has become the default way to add
large-language-model (LLM) features to software that was never designed for them, and such software is almost by
definition legacy: a working REST API or service-layered application that is costly to destabilize~\cite{bisbal1999legacy,khadka2014legacy}.
The practical question is how to add RAG, and the options span a wide intrusiveness spectrum, from a
co-located helper process that never touches legacy source to a single HTTP call written into the nearest handler.
The architecture-erosion literature argues qualitatively that uncontrolled ad hoc integration erodes a system's
design~\cite{vangurp2002erosion}, and work on engineering RAG systems catalogues how RAG systems fail~\cite{barnett2024failure}.
We found no work that measures how intrusive a RAG \emph{retrofit} is, or whether a pattern advertised as
non-intrusive stays so under load and failure. This matters for Industry~5.0, which asks for sustainable and
resilient industry~\cite{breque2021industry50}: adding AI to long-lived industrial software without replacing or
destabilizing it extends the life of existing systems, but only if an AI fault cannot propagate into the operations
that depend on them.

This paper makes four contributions. (1) A \emph{taxonomy} of three retrofit patterns: sidecar (no legacy-code
touch)~\cite{burns2016design,azure2024sidecar,fowler2004strangler}, SDK-embed (a thin client imported into legacy
code, acting as an anti-corruption layer~\cite{evans2003ddd,azure2024acl}), and naive-invasive (an inline HTTP call).
(2) The \emph{Legacy Intrusion Score} (LIS), with three code sub-metrics drawing on the churn and coupling
literature~\cite{nagappan2005churn,chidamber1994metrics}, and \emph{LIS-X}, which adds three operational sub-metrics (run-time overhead, fault propagation, silent feature degradation).
(3) A \emph{static evaluation} on five systems, two of them real codebases downloaded from GitHub. (4) A \emph{live case study} that deploys all
three patterns side by side, each service as its own OS process, and measures latency, staleness, isolation, overhead,
fault propagation, retrieval quality and the swap of four LLM backends.

We ask seven research questions. \textbf{RQ1} Does LIS rank the patterns as their definitions predict, and by how much?
\textbf{RQ2} Does the ranking depend on the weights, or follow from the sub-metrics themselves?
\textbf{RQ3} What latency does the sidecar's RAG path add? \textbf{RQ4} How stale is data behind the sidecar's polling adapter?
\textbf{RQ5} Does load on the RAG layer slow the legacy system on a shared host?
\textbf{RQ6} With all three patterns deployed, what does each cost the legacy request path, and how far does a RAG-layer fault propagate?
\textbf{RQ7} Is the pipeline good enough, in retrieval and answer quality, for these measurements to describe a working system, and can the LLM backend be swapped without touching other components?

\section{Related Work}
\label{sec:related}

\textbf{Legacy systems and erosion.} Bisbal et al.\ survey the issues and migration directions for legacy information systems~\cite{bisbal1999legacy}.
Khadka et al.\ interviewed 26 practitioners and validated the findings with a survey of 198 respondents: practitioners value their
legacy systems highly and the challenges they face are not only technical~\cite{khadka2014legacy}. Van Gurp and Bosch analyse the causes of
design erosion~\cite{vangurp2002erosion}.

\textbf{Integration patterns.} Fowler's Strangler Fig grows a new system around the edges of the old~\cite{fowler2004strangler}.
Evans' anti-corruption layer translates between a new subsystem and a legacy one~\cite{evans2003ddd}; the Azure Architecture Center describes it as a facade that adds latency
and can become a single point of failure, so retry, circuit breaker and bulkhead should be planned~\cite{azure2024acl}. The sidecar
deploys helper components as a separate process or container~\cite{azure2024sidecar,burns2016design}. Our adapter's timestamp-cursor polling is
change-data-capture-style synchronization~\cite{kleppmann2017designing}, not log-based capture.

\textbf{Change metrics.} Nagappan and Ball showed that relative churn measures, which normalize churned and deleted
lines and the number of files churned by component size, discriminate fault-prone from non-fault-prone binaries in Windows Server~2003
with 89.0\% accuracy~\cite{nagappan2005churn}; our churn and breadth follow that lineage, and coupling draws on the Chidamber--Kemerer suite~\cite{chidamber1994metrics}.

\textbf{RAG and resilience.} RAG~\cite{lewis2020rag,gao2024ragsurvey} and the engineering study of RAG failure points~\cite{barnett2024failure}
concern building RAG systems, not retrofitting them. For run-time behaviour we take timeouts, circuit breakers and bulkheads~\cite{nygard2018release} as the yardstick for what a retrofit should contain,
test it with chaos-engineering-style fault injection~\cite{basiri2016chaos}, and follow case-study guidance for the statistics~\cite{runeson2009case,wohlin2012experimentation}.

\section{Patterns and the Legacy Intrusion Score}
\label{sec:lis}

\subsection{Retrofit Patterns}
\textbf{Sidecar.} A co-located, independently deployed process polls the legacy system's existing, unmodified read API and keeps its
own index; no legacy file changes. \textbf{SDK-embed.} Legacy code imports a thin client (\texttt{rag\_client}) that lives in a new file; the legacy handler
calls one function. \textbf{Naive-invasive.} An ad hoc HTTP call is written directly inside a handler, repository method or hook. For every system we
build two naive variants at different layers; they are manually constructed by one researcher, so they capture variance in placement, not between engineers.

\subsection{Code-level LIS}
Let $L$ be the set of a system's original files. For file $f$ let $C_f=|A_f|+|D_f|+|M_f|$ count the added, deleted and modified non-blank lines of a
variant (a replaced line counts once, as modified), and $T_f=|I'_f\setminus I_f|+|E'_f\setminus E_f|$ the new dependency points, where $I_f,E_f$ are the sets of imported modules and
network-endpoint string literals found with Python's \texttt{ast} module before the change and $I'_f,E'_f$ after it.
\begin{align}
\text{churn} &= \frac{\sum_{f\in L} C_f}{\mathrm{LOC}(L)}, \label{eq:churn}\\
\text{breadth} &= \frac{\left|\{f\in L:\ C_f>0\}\right|}{|L|}, \label{eq:breadth}\\
\text{coupling} &= \frac{\sum_{f\in L} T_f}{|L|}, \label{eq:coupling}\\
\text{LIS} &= w_c\,\text{churn}+w_b\,\text{breadth}+w_k\,\kappa, \label{eq:lis}
\end{align}
where $\kappa=\min(\text{coupling},1)$ is the clipped coupling and the default weights are $(w_c,w_b,w_k)=(0.5,0.3,0.2)$. New files (the SDK client, the sidecar adapter) are reported but excluded from churn and breadth; a
legacy file's import of them \emph{is} coupling. Because the sidecar is also scored by this diff (its variant contains only a new file), its LIS of zero is
measured as ``no legacy file touched'', not assigned.

\subsection{LIS-X: operational sub-metrics}
LIS ignores how a retrofit behaves at run time. LIS-X adds three sub-metrics, each in $[0,1]$, measured on the live system of Section~\ref{sec:setup}.
Let $\bar t_{\mathrm{ctrl}}$ and $\bar t_{\mathrm{pat}}$ be the mean latency of the legacy endpoint in the unmodified control and under a pattern.
For each injected fault mode $m$ in $M=\{\text{error},\text{delay}\}$, let $v_m$ be the share of legacy requests that fail or exceed the SLO
$\max(50\,\text{ms},\,3\times\text{control p95})$, and $s_m$ the share of responses that succeed (HTTP status below 400) but lack the AI field; $s_0$ is the same share without a fault.
\begin{align}
\text{runtime} &= 1-\bar t_{\mathrm{ctrl}}/\bar t_{\mathrm{pat}}, \label{eq:runtime}\\
\text{fault} &= \frac{1}{|M|}\sum_{m\in M} v_m, \label{eq:fault}\\
\text{degradation} &= \frac{1}{|M|}\sum_{m\in M}\max\left(0,\ s_m-s_0\right), \label{eq:degr}\\
\text{LIS-X} &= \sum_{i=1}^{6} u_i\,x_i, \label{eq:lisx}
\end{align}
where $x_1,\dots,x_6$ are churn, breadth, $\kappa$, runtime, fault and degradation, and the weights are $u=(0.25, 0.15, 0.10, 0.20, 0.15, 0.15)$.
The runtime term counts only if the bootstrap CI of the latency difference to the control excludes zero. \emph{Fault} records loud damage (failure or slowness) and \emph{degradation} silent damage:
under a fast failure of the RAG layer, a retrofit that swallows errors scores 0 on fault and 1 on degradation; under a delay it is the other way round.
For the control and the sidecar, degradation is 0 by construction, because their legacy endpoint carries no AI field that could disappear; the sidecar's own feature fails visibly, which Table~\ref{tab:ops} reports.
LIS-X exists only where a live deployment exists (System~A).

\textbf{How to read RQ1 and RQ2.} The sidecar touches no legacy file, so its code-level LIS is zero by definition, and one import plus one call is smaller than an inline
call by construction. RQ1/RQ2 therefore check that LIS \emph{discriminates} the patterns and how large the gaps are (construct validity); the falsifiable questions are RQ3--RQ7.

\section{Experimental Setup}
\label{sec:setup}

\begin{figure*}[t]
  \centering
  \resizebox{0.78\textwidth}{!}{\begin{tikzpicture}[
  font=\scriptsize,
  box/.style={rectangle, draw, thick, rounded corners=1pt, align=center,
              text width=2.05cm, minimum height=0.58cm, inner sep=2pt, fill=blue!5},
  smallbox/.style={rectangle, draw, thick, rounded corners=1pt, align=center,
              text width=1.72cm, minimum height=0.58cm, inner sep=2pt, fill=orange!8},
  outbox/.style={rectangle, draw, thick, rounded corners=1pt, align=center,
              text width=1.95cm, minimum height=0.58cm, inner sep=2pt, fill=green!10},
  lane/.style={font=\scriptsize\bfseries},
  arr/.style={-{Latex[length=2mm]}, thick},
  node distance=5mm
]

\node[box]      (corpus)   {5 legacy systems\\A--E (3 synthetic\\+ 2 real, credited)};
\node[box, right=of corpus]      (patterns)  {3 retrofit patterns\\sidecar / SDK-embed /\\naive-invasive $\times2$};
\node[box, right=of patterns]    (diff)      {Diff-based\\measurement:\\churn, breadth, coupling};
\node[box, right=of diff]        (lis)       {Composite LIS\\+ 5 weight vectors\\(sensitivity)};
\node[outbox, right=of lis]      (rq12)      {RQ1 ranking check\\RQ2 sensitivity\\(Table I, Fig. 2)};

\node[lane, above=0.5mm of corpus, anchor=south west, xshift=-2pt] {Static Study (RQ1--RQ2)};

\draw[arr] (corpus) -- (patterns);
\draw[arr] (patterns) -- (diff);
\draw[arr] (diff) -- (lis);
\draw[arr] (lis) -- (rq12);

\node[smallbox, below=11mm of corpus, xshift=-6.9mm]  (legacy)    {Legacy\\monolith\\:5001};
\node[smallbox, right=of legacy]    (adapter)   {ACL / Adapter\\CDC poll, 2s};
\node[smallbox, right=of adapter]   (retrieval) {Retrieval\\MiniLM + FAISS\\:5002};
\node[smallbox, right=of retrieval] (orch)      {Orchestrator\\Groq pool + local\\FLAN-T5 :5003};
\node[smallbox, right=of orch]      (gw)        {Gateway\\:5000};
\node[outbox, right=of gw]          (rq345)     {RQ3 latency (M-W U)\\RQ4 staleness\\RQ5 isolation + \\cross-model};

\node[lane, above=1mm of legacy, anchor=south west, xshift=-2pt] {Dynamic Case Study on System A (RQ3--RQ5)};

\draw[arr] (legacy) -- (adapter);
\draw[arr] (adapter) -- (retrieval);
\draw[arr] (retrieval) -- (orch);
\draw[arr] (orch) -- (gw);
\draw[arr] (gw) -- (rq345);

\begin{scope}[on background layer]
\node[draw=gray!50, dashed, rounded corners=4pt, fit=(corpus)(rq12), inner sep=8pt] {};
\node[draw=gray!50, dashed, rounded corners=4pt, fit=(legacy)(rq345), inner sep=9pt] {};
\end{scope}

\end{tikzpicture}}
  \caption{Research pipeline. \emph{Top:} the static study diffs four variants of each of five legacy systems and computes LIS. \emph{Bottom:} the dynamic study
  deploys System~A four ways (control, sidecar, SDK-embed, naive-invasive) next to the adapter, retrieval, orchestrator and gateway; every box is a separate
  OS process.}
  \label{fig:pipeline}
\end{figure*}
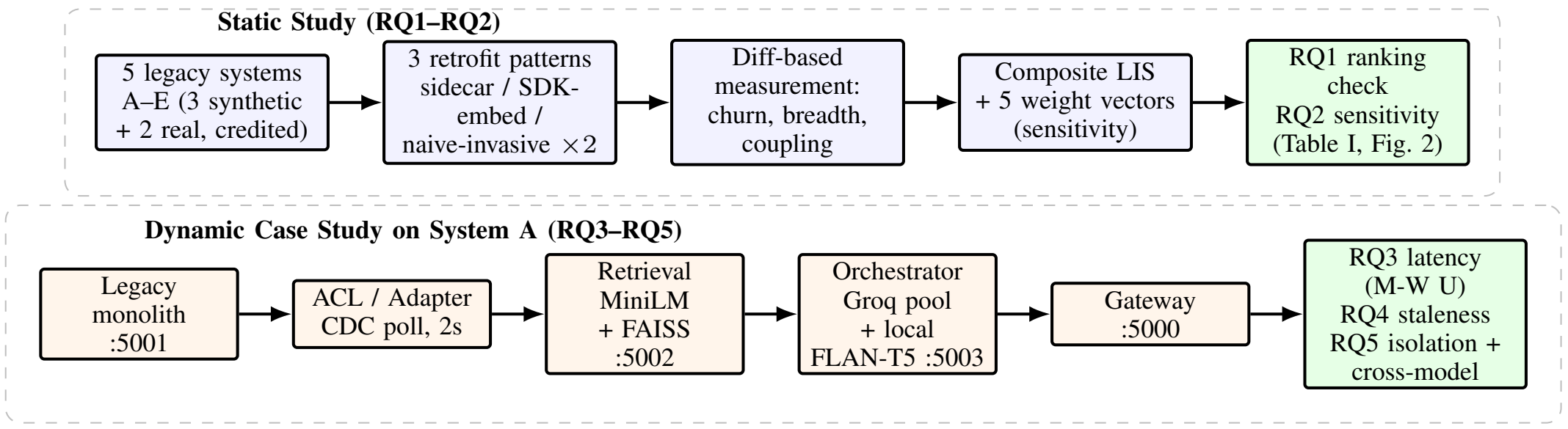

\subsection{Corpus and Variants}
Table~\ref{tab:lis} lists the corpus sizes. Systems A--C are synthetic Python/Flask applications written for the study as multiple files so that breadth is meaningful: a single-layer REST service over SQLite (A),
a repository/service-layered ticket system (B) and an ingest/report batch system (C); B and C place one naive variant in a lower layer (repository, ingestion path), a layering violation. Systems D and E are real, independently authored
codebases that the notebook \emph{downloads} from GitHub at pinned commits (49943b5 and 4b95fb2) and measures verbatim: Flask's tutorial application
\texttt{flaskr}~\cite{pallets2026flask} (BSD-3-Clause; app factory, blueprints, sessions), whose naive variants sit in the \texttt{before\_app\_request} hook that runs on every request
and in the blog index view, and the RealWorld~\cite{realworldspec} Flask backend~\cite{gothinkster2016realworld} (MIT; app factory, dependency injection, JWT, ORM mixin), whose variants sit in a view and in
\texttt{CRUDMixin.save()}, which every model calls on every write. System~E is a three-file excerpt of a larger repository, so its LIS denominators depend on
that choice. Variants of D and E are anchor-checked patches over the downloaded text; if upstream changes so that an anchor stops matching, the build fails.

\subsection{Dynamic Deployment}
The live study runs System~A only, as eight OS processes started with \texttt{subprocess} on one Kaggle host (4 vCPU, Python 3.13.18) (Fig.~\ref{fig:pipeline}).
The four legacy deployments are written from the exact files that the static diff measured, and an assertion checks that recomputing LIS from them reproduces the
static numbers. They share one SQLite file. The adapter polls the sidecar deployment's unmodified \texttt{GET /api/articles} on an \texttt{updated\_at} cursor every
2~s and posts documents to the retrieval service, which embeds them with all-MiniLM-L6-v2~\cite{reimers2019sbert} and indexes them in a FAISS
flat inner-product index~\cite{johnson2019faiss}. The orchestrator is the only component that talks to an LLM; the model is a request field. It serves three models hosted on Groq~\cite{groqdocs}, selected at run time from the live catalogue
(\texttt{gpt-oss-120b} and \texttt{gpt-oss-20b}~\cite{openai2025gptoss}, Qwen3.8-27B~\cite{qwen2026qwen38}), and the local \texttt{flan-t5-small}~\cite{raffel2020t5,chung2022flant5}
through Hugging Face \texttt{transformers}~\cite{wolf2020transformers}; \texttt{gpt-oss-120b} is the default for RQ3 and RQ6. It also hosts a fault-injection switch (error or added delay). The knowledge base has 20 articles; the quality sets are 40 retrieval queries (two phrasings per article) and 12 questions whose answers contain one checkable fact.

\subsection{Protocol and Statistics}
All comparisons use two-sided Mann--Whitney $U$~\cite{mannwhitney1947}, rank-biserial effect size~\cite{kerby2014simple} and percentile-bootstrap 95\% CIs (2,000 resamples)~\cite{efron1979bootstrap}.
RQ3: 50 requests per path after 5 discarded warm-up requests, interleaved in random order. Every request in RQ3--RQ6 is checked for HTTP success (status below 400); latency statistics use successful requests only, and failures would be counted and reported.
RQ4: 40 probes, each at a random phase of the poll cycle (back-to-back probes would be phase-locked to it). RQ5: 100 probes per condition, idle $\to$ load $\to$ idle, with 3 load threads on the CPU-bound local model, probing the legacy
service directly and through the gateway. RQ6: 40 requests per path (interleaved) and 10 per path and fault mode (error; 2~s delay). RQ7: hit@$k$ and MRR against a TF-IDF
baseline~\cite{salton1988tfidf,pedregosa2011sklearn}; answers are scored on the one gold fact after number words are converted to digits (\emph{strict}: the value with its unit within 60 characters, or a bare short reply; \emph{lenient}: the value anywhere),
and every answer that is not strict-correct is reviewed by hand. Backend swap: 3 questions $\times$ 3 repeats per backend through the gateway, changing only the request's model field.

\section{Results}
\label{sec:results}


\begin{table*}[htbp]
\caption{Corpus size (files, non-blank LOC) and LIS sub-metrics (churn, breadth, coupling) with composite LIS by system and pattern (default weights). The sidecar is zero in every row (no legacy file touched) and is omitted.}
\label{tab:lis}
\begin{center}
\footnotesize
\renewcommand{\arraystretch}{1} 
\setlength{\tabcolsep}{3.2pt}
\begin{tabular}{|c|c|c|c|c|c|c|c|c|c|c|c|c|c|c|}
\hline
& & & \multicolumn{4}{c|}{\textbf{SDK-embed}} 
& \multicolumn{4}{c|}{\textbf{Naive-invasive v1}} 
& \multicolumn{4}{c|}{\textbf{Naive-invasive v2}} \\
\cline{4-15}
\textbf{Sys.} & \textbf{Files} & \textbf{LOC} 
& \textbf{Churn} & \textbf{Breadth} & \textbf{Coupl.} & \textbf{LIS}
& \textbf{Churn} & \textbf{Breadth} & \textbf{Coupl.} & \textbf{LIS}
& \textbf{Churn} & \textbf{Breadth} & \textbf{Coupl.} & \textbf{LIS} \\
\hline
A & 2 & 53 
& 0.038 & 0.500 & 0.500 & \textbf{0.269}
& 0.132 & 0.500 & 1.000 & \textbf{0.416}
& 0.132 & 0.500 & 1.000 & \textbf{0.416} \\
\hline
B & 3 & 64 
& 0.047 & 0.333 & 0.333 & \textbf{0.190}
& 0.141 & 0.333 & 0.667 & \textbf{0.304}
& 0.078 & 0.333 & 0.667 & \textbf{0.272} \\
\hline
C & 2 & 36 
& 0.111 & 0.500 & 0.500 & \textbf{0.306}
& 0.250 & 0.500 & 1.000 & \textbf{0.475}
& 0.167 & 0.500 & 1.000 & \textbf{0.433} \\
\hline
D & 4 & 266 
& 0.011 & 0.250 & 0.250 & \textbf{0.131}
& 0.030 & 0.250 & 0.500 & \textbf{0.190}
& 0.023 & 0.250 & 0.500 & \textbf{0.186} \\
\hline
E & 3 & 157 
& 0.019 & 0.333 & 0.333 & \textbf{0.176}
& 0.045 & 0.333 & 0.667 & \textbf{0.256}
& 0.038 & 0.333 & 0.667 & \textbf{0.252} \\
\hline
\end{tabular}
\end{center}
\end{table*}

\subsection{RQ1: Metric Discrimination}
Table~\ref{tab:lis} reports the results. The sidecar's measured LIS is zero in all five systems, and LIS$_{\text{sidecar}}<$ LIS$_{\text{SDK-embed}}<\min$(LIS$_{\text{naive v1}}$, LIS$_{\text{naive v2}}$) in 5/5 systems,
including both real codebases. In System~A the SDK-embed scores 0.269 against 0.416 for the naive variants, driven by churn and coupling (one import is one dependency, an inline call to a service is two:
the import and the endpoint). The real systems have lower absolute LIS than the synthetic ones in every pattern, because the same edit is a smaller fraction of a larger codebase, so absolute values are
not comparable across codebases. No variant deletes code, so deleted lines contribute nothing here.

\subsection{RQ2: Robustness and Dominance}
The ranking holds in 25/25 (system, weight-vector) combinations for five hand-picked weightings and in all systems for 2,000 random weight vectors sampled uniformly from the simplex.
This is expected: in every system each sub-metric of the sidecar is $\le$ that of SDK-embed, which is $\le$ each naive variant's, with a strict improvement in at least one, so the ordering holds for \emph{any}
strictly positive weighting. RQ2 is a dominance property of the sub-metrics, not independent evidence.

\subsection{RQ3: Latency Overhead}
Through the live gateway the legacy-only path takes 7.6~ms on average and the RAG-augmented path, served by \texttt{gpt-oss-120b}, 348.0~ms (median 327, p95 433, range 175--1423~ms): a mean overhead of 340.4~ms (95\% CI $[303.2,\,393.0]$), $U{=}0$, $p{=}7.1\times10^{-18}$,
$r{=}1.00$ ($N{=}50$ each; all 100/100 requests returned HTTP 200). The RAG path includes retrieval, an external LLM call and two extra service hops, and its latency depends on the model (209--686~ms across the four backends in RQ7),
so the overhead is specific to this configuration and is not a general RAG constant.

\subsection{RQ4: Staleness}
39 of 40 probes became retrievable. The remaining probe had been indexed but ranked 12th of 43 documents, outside the top-10 criterion for retrievability, so we count it as failed and exclude it.
Server-side staleness (legacy write timestamp to index insertion, which excludes the probe loop's own polling) has mean 911~ms (95\% CI $[730,\,1093]$), median 908, p95 1969 and maximum 2026~ms (Fig.~\ref{fig:results}c).
This matches writes landing uniformly within a 2~s poll cycle plus a few tens of milliseconds of indexing; the probe loop added a mean detection delay of 72~ms.

\subsection{RQ5: Runtime Isolation}
With every service a separate OS process there is no shared interpreter, yet load on the RAG layer slowed the legacy service from 3.63 to 5.47~ms when probed directly (1.51$\times$, 95\% CI $[1.45,\,1.57]$; $r{=}0.98$) and from 6.99 to 11.00~ms through the gateway (1.57$\times$, CI $[1.52,\,1.64]$).
The idle block after the load matches the one before it (3.61 vs.\ 3.63~ms directly, 1.51$\times$ against either), so drift does not explain the effect. All 100 probes in each of the six blocks
succeeded, and 6 load requests completed (0 failed). The absolute added delay is about 2--4~ms, but it shows that process separation on one host does not isolate resources; the loaded block is short, so this is a lower-bound signal.


\begin{table*}[htbp]
\caption{RQ6: legacy endpoint \texttt{GET /api/articles/<id>} per deployment and under injected RAG-layer faults (ms unless noted; Fail = HTTP error or exception; No AI = response lacks the AI field; SLO = share of requests failing or slower than 50~ms).}
\label{tab:ops}
\begin{center}
\footnotesize
\renewcommand{\arraystretch}{1}
\setlength{\tabcolsep}{4pt}
\begin{tabular}{|l|c|c|c|c|c|c|c|}
\hline
& \multicolumn{2}{c|}{\textbf{No fault}}
& \multicolumn{3}{c|}{\textbf{Fault: orchestrator returns error}}
& \multicolumn{2}{c|}{\textbf{Fault: 2~s delay}} \\
\cline{2-8}
\textbf{Path} & \textbf{Mean} & \textbf{p95}
& \textbf{Mean} & \textbf{Fail \%} & \textbf{No AI \%}
& \textbf{Mean} & \textbf{SLO viol. \%} \\
\hline
Control (no retrofit) & 3.8 & 4.5 & 3.7 & 0 & -- & 3.7 & 0 \\
\hline
Sidecar & 3.9 & 4.3 & 3.8 & 0 & -- & 3.8 & 0 \\
\hline
SDK-embed & 572.3 & 2383.0 & 7.2 & 0 & 100 & 2440.8 & 100 \\
\hline
Naive-invasive & 479.1 & 504.0 & 7.2 & 0 & 100 & 2426.5 & 100 \\
\hline
Sidecar RAG feature (\texttt{/ask}) & 618.9 & 2383.1 & 7.0 & 100 & -- & 2467.7 & 100 \\
\hline
\end{tabular}
\end{center}
\end{table*}

\subsection{RQ6: Operational Intrusion of All Three Patterns}
\textbf{Run-time overhead.} Against the unmodified control (3.80~ms) the sidecar deployment shows no detectable change (difference 0.07~ms, CI $[-0.10,\,0.24]$, $p{=}0.31$), although the adapter polls it
every 2~s. SDK-embed and naive-invasive add 569~ms (CI $[418,\,752]$) and 475~ms (CI $[413,\,583]$), both $p<10^{-13}$, because the LLM call now runs inside the legacy handler.
The two retrofits are not statistically distinguishable from each other ($p{=}0.62$; difference 93~ms, CI $[-97,\,304]$).
The sidecar's own feature path is not distinguishable from the SDK path ($p{=}0.13$); against the naive path $p{=}0.029$ without correction for three comparisons. Medians are similar (397, 427 and 436~ms for feature, SDK and naive) and the tails heavy (4, 3 and 1 of 40 requests took 2.3--3.5~s),
so in the typical case the feature costs about the same under the three patterns; the difference is where the cost is incurred, which for the two retrofits is the legacy endpoint.
All 200/200 requests succeeded.

\textbf{Fault propagation (Table~\ref{tab:ops}).} Two kinds of damage appear. \emph{Loud:} under a 2~s delay, the SDK and naive endpoints average 2.44 and 2.43~s with 100\% SLO violations, while the sidecar's legacy endpoint stays at 3.8~ms.
\emph{Silent:} when the orchestrator fails fast, the SDK and naive endpoints still answer with HTTP success, within the SLO (their latency rises from 3.7 to 7.2 and 7.2~ms), but the AI field is absent from 100\% of responses and no error is raised,
whereas the sidecar's feature returns errors in 100\% of requests and its legacy endpoint is unaffected.
Both retrofits call the orchestrator synchronously and swallow errors; a fixed 10~s timeout is their only stability measure, with no circuit breaker, bulkhead or asynchronous execution~\cite{nygard2018release}.


\begin{table}[htbp]
\caption{LIS-X sub-metrics for System~A (churn, breadth and coupling as in Table~\ref{tab:lis}; LIS is the code-only composite).}
\label{tab:lisx}
\begin{center}
\footnotesize
\renewcommand{\arraystretch}{1}
\setlength{\tabcolsep}{3pt}
\begin{tabular}{|l|c|c|c|c|c|c|c|c|}
\hline
\textbf{Pattern} & \textbf{Churn} & \textbf{Bread.} & \textbf{Coupl.}
& \textbf{Runt.} & \textbf{Fault} & \textbf{Degr.}
& \textbf{LIS} & \textbf{LIS-X} \\
\hline
Sidecar & 0.000 & 0.000 & 0.000 & 0.000 & 0.000 & 0.000 & 0.000 & \textbf{0.000} \\
\hline
SDK & 0.038 & 0.500 & 0.500 & 0.993 & 0.500 & 0.500 & 0.269 & \textbf{0.483} \\
\hline
Naive & 0.132 & 0.500 & 1.000 & 0.992 & 0.500 & 0.500 & 0.416 & \textbf{0.556} \\
\hline
\end{tabular}
\end{center}
\end{table}

\textbf{LIS-X.} The sidecar scores 0.000, SDK-embed 0.483 and naive-invasive 0.556 (Table~\ref{tab:lisx}). The operational terms are nearly equal for the two retrofits (runtime 0.993 vs.\ 0.992, fault 0.50 and degradation 0.50 for both), so they erase half of SDK-embed's code-level advantage
(the SDK--naive gap falls from 0.147 to 0.073), and the ranking holds in 100\% of 2,000 random six-dimensional weight vectors, but only because the code terms still favour SDK-embed.

\subsection{RQ7: Retrieval, Answer Quality and Backend Swap}
\begin{figure*}[t]
  \centering
  \includegraphics[width=\textwidth,height=7.7cm]{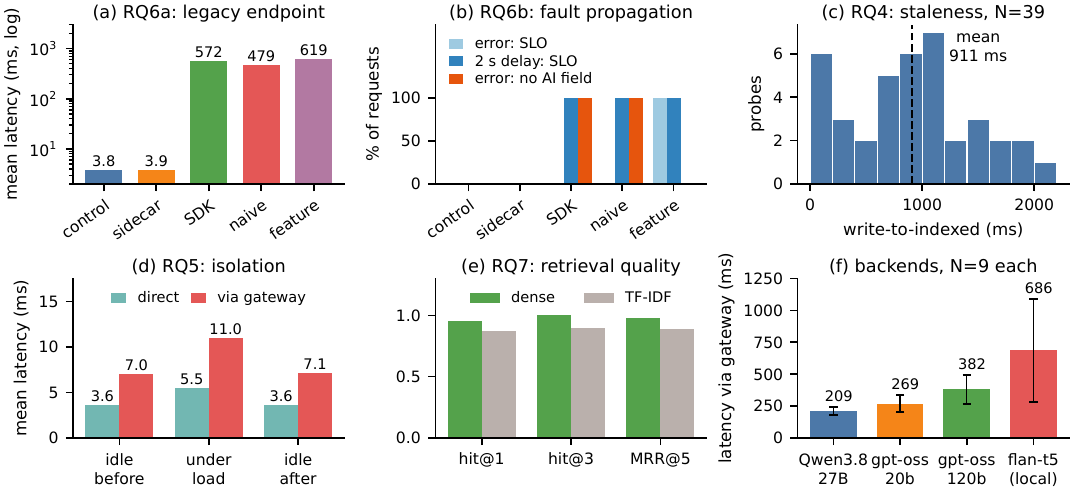}
  \caption{Dynamic results on System~A. (a) Mean latency of the legacy endpoint per deployment (log scale; ``feature'' is the sidecar's RAG path). (b) Share of requests violating the SLO under each fault, and share of responses without the AI field (silent degradation). (c) Server-side staleness. (d) Legacy latency before, during and after load on the RAG layer.
  (e) Retrieval quality. (f) Mean latency (sd) per LLM backend through the full pipeline.}
  \label{fig:results}
\end{figure*}

On 40 queries over 20 documents, dense retrieval reaches hit@1 0.95 (38/40, bootstrap CI $[0.875,\,1.000]$), hit@3 1.00 and MRR@5 0.975, against 0.875, 0.90 and 0.888 for TF-IDF;
dense retrieval ranks the gold article first on 3 queries where TF-IDF does not, and TF-IDF on 0 where dense retrieval does not.
Each hosted model answers all 12 questions correctly and \texttt{flan-t5-small} 9/12; strict and lenient scoring agree on all 48 answers. Reading all 48 confirmed that every match is correct and that the 3 flagged answers, all from \texttt{flan-t5-small}, are wrong:
two reproduce a different article (a travel question answered with the meeting-room article, a remote-work question with the leave carry-over article) and one is ``unanswerable''. The gold article was retrieved in all 48/48 cases, so these are generation errors.
Across the four backends, all 36/36 invocations through gateway and orchestrator succeeded with only the request's model field changed, with mean latencies of 209 (Qwen3.8-27B), 269 (gpt-oss-20b), 382 (gpt-oss-120b) and 686~ms (flan-t5-small; Fig.~\ref{fig:results}f).
The surrounding code is unchanged, so the latency differences come from the backends. This is a sanity check, not a benchmark.

\section{Discussion}
\label{sec:discussion}
Three points follow. First, code-level cleanliness is not run-time containment: SDK-embed wins on every code dimension yet matches the naive call at run time, so a retrofit should be judged on both. An SDK is the natural place to put
timeouts, a circuit breaker and asynchronous execution~\cite{nygard2018release}; the SDK evaluated here does not, which its operational scores reflect. Second, swallowing errors turns a loud failure into a silent one: the legacy endpoint stays fast and returns success
while the feature is gone, so legacy health checks cannot see it and feature-level success must be monitored. Third, the sidecar's isolation costs staleness of about half the polling interval and an extra process, and it is not free even across processes on one host, so ``zero coupling'' is a code-level claim.

\section{Threats to Validity}
\label{sec:threats}
\textbf{Construct.} LIS and LIS-X are proxies; weights and the SLO threshold are choices, and RQ1/RQ2 are near-definitional (Section~\ref{sec:lis}). Degradation is defined only for the two retrofits whose legacy endpoint should carry the AI field. Answer correctness checks one gold fact per question; the strict scorer and the manual review of every answer limit, but do not remove, errors in either direction.
\textbf{Internal.} All services share one host (4 vCPU), so contention is host-level. The hosted models are external services whose latency varies and whose catalogue changes, so the model IDs are recorded; the live study is a single run without replicates, and the RQ5 loaded block is short. Interleaving, idle--load--idle and random probe phase reduce, but cannot remove, order effects.
\textbf{External.} The corpus is small, Python/Flask-only, and System~E is a three-file excerpt; the dynamic study covers System~A only. All variants were written by one researcher, and ``naive'' as implemented (synchronous, 10~s timeout, errors swallowed)
may not match real code; an asynchronous SDK would score differently on RQ6.
\textbf{Conclusion.} Sample sizes are modest (10 to 100 observations per condition; 12 answer questions; 9 backend invocations per model), and the corpus supports only the claim that the ranking held in the systems tested.

\section{Conclusion and Future Work}
\label{sec:conclusion}
We presented a taxonomy of RAG-retrofit patterns, the code-level LIS, and LIS-X with run-time overhead, fault propagation and silent feature degradation. Statically, the sidecar $<$ SDK-embed $<$ naive-invasive ranking held in five systems and under every weighting, but only because
of dominance. Live, the sidecar left the legacy endpoint untouched while both in-process retrofits added hundreds of milliseconds, passed a dependency delay through, and lost the feature silently when the RAG layer failed fast; SDK-embed gave code-level but not run-time containment. Four LLM backends were swapped by changing one request field. Future work
should (i) have independent engineers write the integrations, (ii) give the SDK timeouts, circuit breaking and asynchronous execution and re-measure, (iii) replicate the live study over several runs and backends, (iv) extend the corpus beyond Flask/Python, and (v) use separate hosts or containers.

\section*{Data and Artifact Availability}
The artifact contains the notebook and its executed copy (static study, service scripts, experiments), the raw per-request measurements and summary tables (CSV/JSON), the script that generates every number and figure in this paper from them,
per-run metadata (versions, model identifiers, configuration) and the pinned upstream commit hashes. Upstream code is downloaded at run time, not redistributed; licence texts are saved with the results.
The notebook stores no credentials: Groq keys are read from Kaggle Secrets, and without them the run falls back to the local model. Hosted-model results vary between runs. The artifact is available at \textit{https://github.com/TahlilMahfuz/Quantifying-the-Intrusiveness-of-Retrofitting-RAG-onto-Legacy-Software-Systems.git}.

\section*{Acknowledgment}
We thank the Pallets Projects (Flask tutorial~\cite{pallets2026flask}, BSD-3-Clause) and the gothinkster/RealWorld contributors (MIT)~\cite{gothinkster2016realworld,realworldspec}, whose code was measured unmodified, and Groq, OpenAI and the Qwen team for the hosted and open-weight models~\cite{groqdocs,openai2025gptoss,qwen2026qwen38}.

\bibliographystyle{IEEEtran}
\bibliography{References_v6}
\end{document}